\documentclass{llncs}

\usepackage{color}
\usepackage{listings}
\usepackage{float}
\usepackage{url}
\usepackage{tikz}
\usepackage{graphicx}
\usepackage{multirow}

\title{
RDF Knowledge Graph Visualization From a Knowledge Extraction System}
\author{
Fadhela Kerdjoudj \inst{1,2}
\and
Olivier Cur\'e \inst{1}
}

\institute{
 Universit\'e Paris-Est Marne-la-vall\'ee, LIGM, CNRS UMR 8049, France.,\\
\and
GEOLSemantics, 12 rue Raspail 94250, Gentilly, France.\\
\email{$\lbrace$fadhela.kerdjoudj ,ocure$\rbrace$@u-pem.fr }
}

\begin{document}

\maketitle

\begin{abstract}
In this paper, we present a system to visualize RDF knowledge graphs. These graphs are obtained from a knowledge extraction system designed by GEOLSemantics. This extraction is performed using natural language processing  and trigger detection. The user can visualize subgraphs by selecting some ontology features like concepts or individuals. The system is also multilingual, with the use of the annotated ontology in English, French, Arabic and Chinese. 
\end{abstract}

\section{Introduction}
During the last decades, many knowledge extraction systems have emerged aiming at automating textual document processing. The importance of these systems is highlighted by the proliferation of textual publications on the web, $e.g.$ social media, blogs or journals. In order to extract as much  relevant information as possible, it is necessary to exploit the power offered by the semantic web and its related technologies. These technologies, namely, vocabularies (RDF, OWL, SKOS...), query language (SPARQL), inference services, Linked Open Data (LOD), allow to represent, access, reason over and interconnect extracted data. To obtain these data, we use a knowledge extraction system based on Natural Language Processing (NLP) to populate a Knowledge Base.

In this article we present a component of this system which deals with  RDF knowledge graph visualization.
It allows to build subgraphs by selecting either ontology concepts or  individuals. 
Indeed, the size of the knowledge graph extracted is proportional to the length of the text. Therefore, the graph could become fairly large and dense. 
To deal with this issue, we propose an approach that helps visualizing and summarizing the extracted knowledge.
\\In this article, we begin by introducing the system used to perform the knowledge extraction and which processing it performs. We then present our approach to visualize the knowledge graph. 

\section{Knowledge extraction system}
The Web contains a huge number of documents from heterogeneous sources. 
However, these documents cannot be used directly by programs.
\\The extraction and representation framework developed at GEOLSemantics, combines NLP techniques with semantic processing  to extract RDF knowledge graphs. In the rest of this section, we describe the main performed tasks.

\subsection{Deep Morphosyntactic Analysis} 
The deep morphosyntactic analysis consists of the following steps.
\begin{itemize}

\item Segmentation:
the text is split into tokens using regular expressions which identify capital letters, numbers, dates, \textit{etc.}
\item Morphological Processing:
for each inflected word the basis (lemma) is identified and a grammatical category is attributed to it. 
\item Named entity recognition:
the named entities, namely: Person, Organization, Location are identified using two methods: \textit{(i)} Thesaurus consultation, based on LOD such as DBpedia and Geonames. \textit{(ii)} Declarative rules based on announcers, such as President, Mister, city, airport. 
\item Syntactic Analysis:
allows to represent the syntactic structure of a text. It indicates how the grammatical categories are arranged, \textit{e.g.}, noun-verb-adjective. 
Indeed, some other processes are performed like: transform passive forms to active, resolve anaphora, detect negation and verb tense which gives information about the modalities of the action.

\end{itemize}
\subsection{Knowledge Extraction}
The knowledge extraction allows to identify entities and relations between them. This is performed using an ontology-based approach which defines the different concepts needed for annotating entries of the original text. 
All these processes are organized as follows:
\begin{itemize}
\item \textit{Probable Concept Selection}:
It consists in spotting triggers. Triggers are composed of one or several words (nouns, verbs, \textit{etc.}) that represent a semantic unit denoting an entity to extract. For instance, the verb ''go" denotes a Displacement.
Each trigger is associated to an ontology class and a number of rules. 
\item \textit{Rule Selection}:
Each trigger is associated to a list of rule patterns. From the different relations identified in the syntactic analysis a matching approach enables to select all relevant patterns. 
\item \textit{Triple Creation}: the rules selected help to create triples using the given patterns. Then the result is structured as RDF triples.
\end{itemize}

\subsection{Integration}
In this step, our aim is to bring more consistency to the extracted knowledge by performing the following processes: \textit{(i) }Coreference resolution: as described in \cite{dredze2010entity}, we group all the instances of each entity. \textit{(ii) }Relative dates resolution: transform all relative dates like \textit{today, last week} to absolute dates.
 \textit{(iii) }Complete the extraction with implicit information which can be inferred when reading a text such as the date or the place. 
 \textit{(iv) }Label creation: following the token positions indicated by the morphosyntactic processing, the labels are retrieved from the original text. It helps to identify each entity. 

\section{Visualization Features}
\subsection{Ontology description}
Using an RDF triple based representation and ontology description, the text can be represented as a knowledge graph which contains all needed information. 
\\Currently our ontology is expressed in OWL (precisely $ACLIF(D)$ description logic) and contains a few hundred classes and properties. We are constantly enriching this ontology to support more use cases and domains of expertise.
The classes considered are mainly: 
\begin{itemize}
\item Named entities such as person, organization, location, measure, date.
\item Facts such as professional experience, studies, family relation, personal relation, event relation, organization relation.
\item Events like meeting, movement, violent act, conviction, appointment, arrest.
\end{itemize}
The object properties describe the relation between entities such as the address of a person or an organization, the date and the place where an event takes place.
Finally the datatype properties are literals which describe the named entities such as names, types, values.   
\\It is worth noticing that a great importance has been given to the ontology design. Indeed all the classes and properties have to be labeled in different languages. At this point, our ontology contains Arabic, French, English and Chinese. Also, all the properties must be related to their respective domain and range. 
\subsection{Graph features}
The extracted knowledge is compliant with the ontology description, the triples are related to each other and the graph can be constructed. In our graph representation, instead of using URIs to denote nodes, we use icons and labels. This allows the user to spot the requested information in an easier way than reading all the URIs which are usually less illustrative. The edges are also denoted with labels as they were indicated in the ontology.
\subsubsection{Multilingual aspect:}
\begin{figure}[!h]
\includegraphics[scale=0.6]{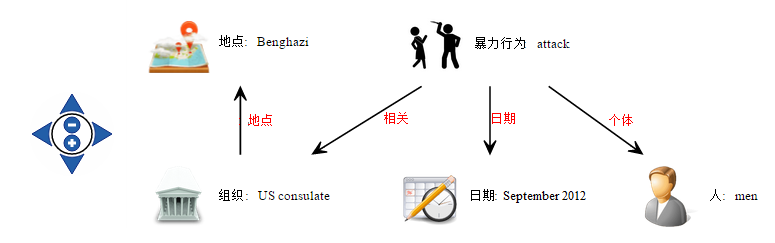}
\caption{
\label{figure_Benghazi_consulat} RDF graph of the example \ref{Ex_Begnghazi_Consulat}}
\end{figure}

Defining the labels in the ontology in different languages allows to visualize the graph in a multilingual form as shown in Figure \ref{figure_Benghazi_consulat} which describes the knowledge extracted from Example \ref{Ex_Begnghazi_Consulat}. By selecting the Chinese language, the user can visualize the knowledge graph corresponding to an English text using Chinese annotation.
We note that literals could also be translated into the selected language. For the sake of clarity, we keep them in their original form (\textit{i.e.} like they were cited in the text).

\begin{example}\label{Ex_Begnghazi_Consulat}
In September 2012, the US consulate in Benghazi was attacked by armed men.
\end{example}
\vspace{-5mm}
\vspace{-7mm}
\subsubsection{Faceted search:}
\begin{figure}[!h]
\includegraphics[scale=0.5]{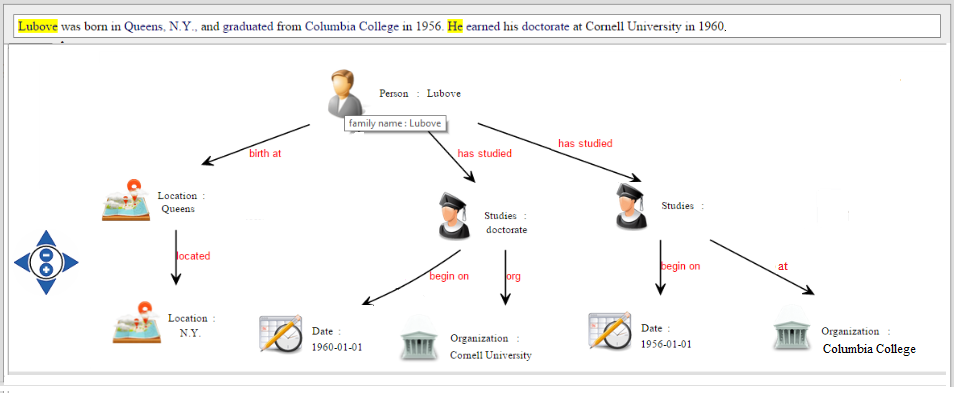}
\caption{
\label{figure_Studies} Graph with tooltip and links to the text.}
\end{figure}

As we have already stated, the graph can be dense and hardly understandable, here we propose a selection of subgraphs which can help the user to directly visualize the relevant information. Two selections are proposed :
\begin{enumerate}
\item \textbf{Concept Selection: }
The RDF is parsed in order to retrieve all the classes instantiated in the viewed text. Hence, we select all the rdf:type that the RDF graph contains. For instance, Person, Location, Organization, ViolentAct and Date in Example \ref{Ex_Begnghazi_Consulat}.
\item \textbf{Individual Selection: }
All individuals extracted from the text are proposed to the user. Labels are used instead of URIs to help the user to select the needed individual. In Example \ref{Ex_Begnghazi_Consulat}, the instances are: Benghazi, attack, man, September 2012, US consulate. 
\end{enumerate}
In the latter, the user can select the graph degree depth. It indicates how deep the subgraph must be, \textit{i.e.} if adjacent nodes need to be developed.\\To avoid cluttering, we only display relations which denote object properties. Datatype properties are viewed when hovering a node as tooltips as shown in Figure \ref{figure_Studies}. 
We also propose a table view of the triples extracted. The first column contains the Subject, the second, the predicate and the third one, the Object. 

\section{Implementation}
The visualization module of our system is a web interface developed in Java. The graph is built using GraphViz\footnote{http://www.graphviz.org/Documentation.php}, where nodes denote instance classes and edges relations between them. Graphviz \cite{ellson2004graphviz} constructs a graph from an entry in DOT \cite{koutsofios1991drawing} language. The diagrams are rendered in different formats: PNG, PDF, SVG, \textit{etc.} The text (labels) can be handled via useful features such as: font, color, size, hyperlinks, custom shapes. In addition, the graph layout can be hierarchical, radial or circular. Moreover, we used some javascript code to link the graph to the text by highlighting the trigger when hoovering the corresponding node. Finally, the ontology is parsed with the Jena API \cite{carroll2004jena}, to retrieve classes and properties, hierarchies and annotations.

\section{Conclusion and future work}
In this work, we present a system to visualize an RDF knowledge graph. It allows to select subgraphs. This feature is especially useful in the case of big graphs obtained from long text processing. We also explain the role played by the ontology in the visualization, it helps providing more clarity and offers a multilingual interpretation of the text.
As future work, we prospect to handle more RDF extractions such as Yago and DBpedia, and make the graph more interactive by allowing to move the nodes for instance. 
\bibliographystyle{plain}
\bibliography{biblioDerive} 	

\begin{thebibliography}{1}

\bibitem{carroll2004jena}
Jeremy~J Carroll, Ian Dickinson, Chris Dollin, Dave Reynolds, Andy Seaborne,
  and Kevin Wilkinson.
\newblock Jena: implementing the semantic web recommendations.
\newblock In {\em Proceedings of the 13th international World Wide Web
  conference on Alternate track papers \& posters}, pages 74--83. ACM, 2004.

\bibitem{dredze2010entity}
Mark Dredze, Paul McNamee, Delip Rao, Adam Gerber, and Tim Finin.
\newblock Entity disambiguation for knowledge base population.
\newblock In {\em Proceedings of the 23rd International Conference on
  Computational Linguistics}, pages 277--285. Association for Computational
  Linguistics, 2010.

\bibitem{ellson2004graphviz}
John Ellson, Emden~R Gansner, Eleftherios Koutsofios, Stephen~C North, and
  Gordon Woodhull.
\newblock Graphviz and dynagraph—static and dynamic graph drawing tools.
\newblock In {\em Graph drawing software}, pages 127--148. Springer, 2004.

\bibitem{koutsofios1991drawing}
Eleftherios Koutsofios, Stephen North, et~al.
\newblock Drawing graphs with dot.
\newblock Technical report, Technical Report 910904-59113-08TM, AT\&T Bell
  Laboratories, Murray Hill, NJ, 1991.

\end{thebibliography}
\end{document}